\documentclass{article}

\usepackage{etoolbox}
\newtoggle{arxiv}
\toggletrue{arxiv} % or \togglefalse{arxiv}

\usepackage{xcolor}
\usepackage{amsmath}

\definecolor{myblue}{rgb}{0.21,0.49,0.74}

\iftoggle{arxiv}{
    \usepackage[bookmarks=false, colorlinks,allcolors=myblue,pdfborder={0 0 0},pdfborderstyle={/S/U/W 0}]{hyperref}
}{
    \usepackage[bookmarks=false,pdfauthor={\authorname},pdfsubject={\pdfsubject},hidelinks]{hyperref}
}

\usepackage{cleveref}

\newcommand\blfootnote[1]{%
  \begingroup
  \renewcommand\thefootnote{}\footnote{#1}%
  \addtocounter{footnote}{-1}%
  \endgroup
}

\newcommand{\magnet}[0]{\texttt{MAGNeT}}
\newcommand{\musicgen}[0]{\texttt{MusicGen}}

\newcommand{\encodec}[0]{\texttt{EnCodec}}

\usepackage{booktabs}
\usepackage{arydshln}

\usepackage[T1]{fontenc}
\usepackage[utf8]{inputenc}

\iftoggle{arxiv}{
    \usepackage{ismir}
}{
    \usepackage[submission]{ismir} % Remove the "submission" option for camera-ready version
    \usepackage{cite}
}

\usepackage{url}
\usepackage{listings}
\usepackage{float}
\usepackage{graphicx}
\usepackage{color}
\usepackage{pifont}
\usepackage{amsmath, amssymb}
\usepackage{linguex}

\definecolor{commentgreen}{rgb}{0.1,0.4,0.1}
\definecolor{keywordgreen}{rgb}{0,0.55,0}
\definecolor{codegray}{rgb}{0.5,0.5,0.5}
\definecolor{codepurple}{rgb}{0.58,0,0.82}
\definecolor{functionblue}{rgb}{0.2,0.2,.8}
\definecolor{additioncolor}{rgb}{0.7,0.3,0}

\lstdefinestyle{mystyle}{
    backgroundcolor=\color{white},   
    commentstyle=\color{commentgreen}\itshape,
    keywordstyle=\color{keywordgreen}\bfseries,
    numberstyle=\tiny\color{codegray},
    stringstyle=\color{codepurple},
    basicstyle=\ttfamily\tiny,
    breakatwhitespace=false,         
    breaklines=true,                 
    captionpos=b,                    
    keepspaces=true,                 
    numbers=left,                    
    numbersep=5pt,                  
    showspaces=false,                
    showstringspaces=false,
    showtabs=false,
    tabsize=2
}

\usepackage{algorithm}
\usepackage{algpseudocode}

\graphicspath{{figures/}}

\makeatletter
\def\adl@drawiv#1#2#3{%
\hskip.5\tabcolsep
\xleaders#3{#2.5\@tempdimb #1{1}#2.5\@tempdimb}%
#2\z@ plus1fil minus1fil\relax
\hskip.5\tabcolsep}
\newcommand{\cdashlinelr}[1]{%
\noalign{\vskip\aboverulesep
\global\let\@dashdrawstore\adl@draw
\global\let\adl@draw\adl@drawiv}
\cdashline{#1}
\noalign{\global\let\adl@draw\@dashdrawstore
\vskip\belowrulesep}}
\makeatother

\definecolor{bg}{gray}{0.95}

\graphicspath{{figures/}}

\title{LoopGen: Training-Free Loopable Music Generation}

\oneauthor
  {Anonymous Authors}
  {Anonymous Affiliations\\\texttt{anonymous@ismir.net}}

\multauthor
  {Davide Marincione$^{\star}$ \hspace{1cm} Giorgio Strano$^{\star}$ \hspace{1cm} Donato Crisostomi}
  {{\bf Roberto Ribuoli \hspace{1cm} Emanuele Rodolà}\\
  Sapienza University of Rome\\
  {\tt\small \{marincione, strano\}@di.uniroma1.it}
  }

\def\authorname{F. Author, S. Author, and T. Author}

\begin{document}

\maketitle

\begin{abstract}
    Loops--short audio segments designed for seamless repetition--are central to many music genres, particularly those rooted in dance and electronic styles.\blfootnote{$\star$ denotes equal contribution.} However, current generative music models struggle to produce truly \emph{loopable} audio, as generating a short waveform alone does not guarantee a smooth transition from its endpoint back to its start, often resulting in audible discontinuities.
    We address this gap by modifying a non-autoregressive model (\magnet{}) to generate tokens in a circular pattern, letting the model attend to the beginning of the audio when creating its ending. This inference-only approach results in generations that are aware of future context and loop naturally, without the need for any additional training or data. We evaluate the consistency of loop transitions by computing token perplexity around the seam of the loop, observing a 55\% improvement. Blind listening tests further confirm significant perceptual gains over baseline methods, improving mean ratings by 70\%. Taken together, these results highlight the effectiveness of inference-only approaches in improving generative models and underscore the advantages of non-autoregressive methods for context-aware music generation.
    \begin{center}
        \footnotesize
        \hspace{0.2cm}\raisebox{-0.2\height} {\includegraphics[width=0.9em,height=0.9em]{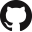}}\hfill \href{https://github.com/gladia-research-group/loopgen}{\texttt{github.com/gladia-research-group/loopgen}}

        \hspace{0.2cm}\raisebox{-0.2\height}{\includegraphics[width=0.9em,height=0.9em]{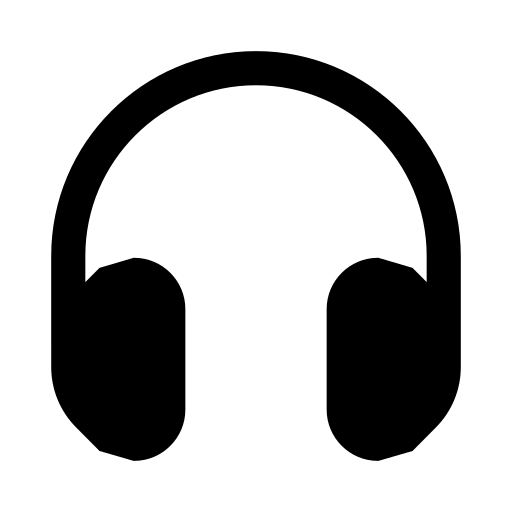}}\hfill \href{https://gladia-research-group.github.io/loopgen-demo/}{\texttt{gladia-research-group.github.io/loopgen-demo}}
    \end{center}
\end{abstract}

\section{Introduction}
\label{sec:introduction}
Loops play a critical role in music production across a broad range of genres, from hip-hop to electronic dance music. By definition, a loop is a segment of audio that can be repeated indefinitely without noticeably jarring transitions between consecutive repetitions. These short segments function as building blocks in many compositions, providing rhythmic and harmonic foundations that can be layered, remixed, and manipulated. Indeed, entire online platforms (e.g., Splice\footnote{\url{https://splice.com/}}) revolve around sharing and curating loops, underscoring their commercial and creative significance in contemporary music-making.

However, despite their ubiquity in practice, loops remain an underexplored challenge for generative music models. The primary issue lies in the disconnect between \emph{generating a short audio sample} and \emph{ensuring that it loops correctly}. Many existing generative approaches focus on producing samples that sound coherent when played from start to finish \cite{musicgen, magnet, jukebox, audioldm, mousai, riffusion}, but they do not explicitly consider the transition point from the end of the sample back to its beginning. As a result, naive repetition of these segments often yields abrupt discontinuities, limiting their practical utility for musicians and producers who rely on seamless repetition.
\begin{figure}
    \centering
    \includegraphics[alt={Diagram showing MAGNeT's input divided into three sections: a central tile, a left padding, and a right padding. The central tile contains the main audio tokens to be generated. The left padding duplicates the tile's rightmost tokens, while the right padding duplicates its leftmost tokens. This structure encourages smoother looping by exposing the model to boundary conditions during inference.},width=\linewidth]{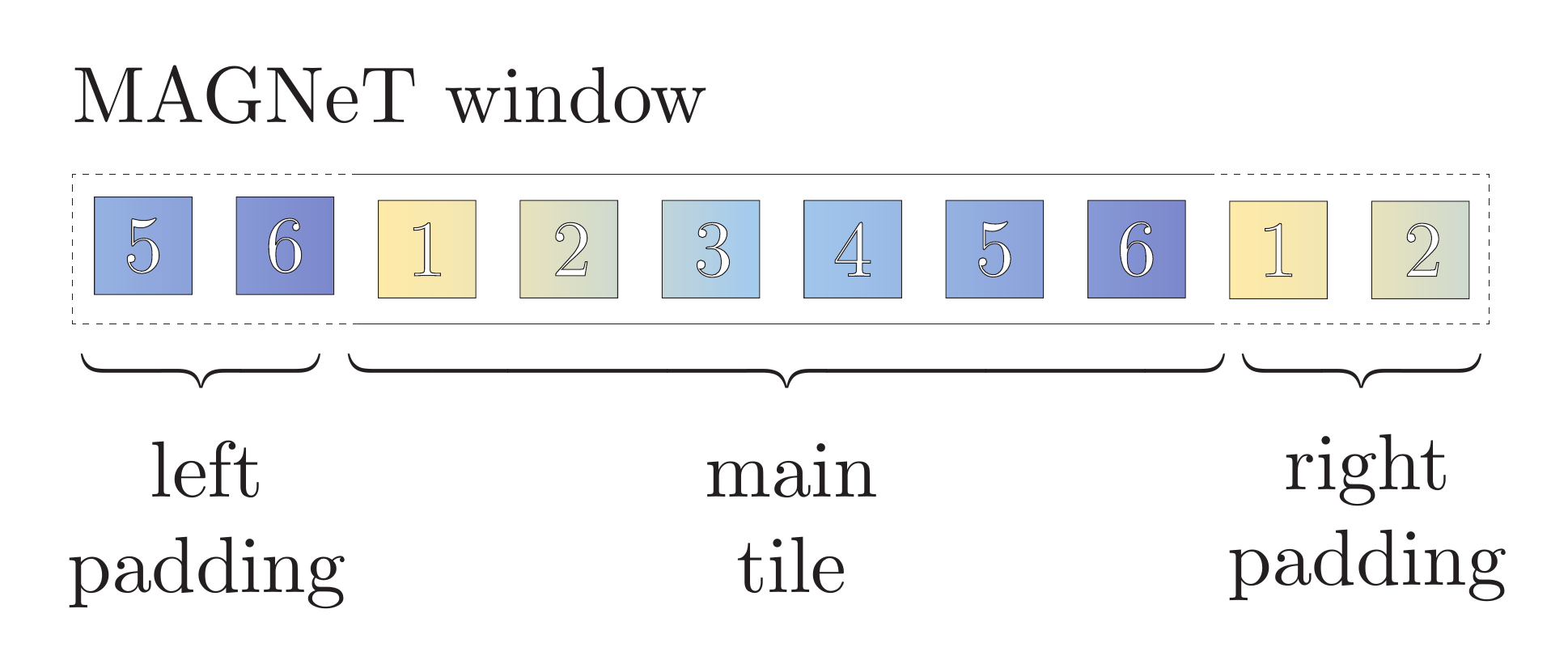}
    \caption{Our proposed circular padding framework for loopable sample generation.}
    \label{fig:teaser}
\end{figure}

In this paper, we introduce a loop-aware generation framework that modifies the \emph{iterative inference} of a non-autoregressive (NAR) model to produce seamless loops. Concretely, we adopt a \emph{circular padding} strategy, replicating partial portions of the loop at both ends of the generation window, so that the model attends to the loop’s beginning while generating its ending (\Cref{fig:teaser}). This ensures a smooth endpoint-to-onset transition, effectively creating “bridging tokens” that align the tail of the sample with its onset. Our method can be used in two ways: (1) to generate entire loopable segments from scratch, or (2) to refine the end of an existing audio sample so that it loops seamlessly. Additionally, we implement a \emph{beat-aware} technique that constrains the total length of the loop to align with musical bars, further promoting coherent repetition.

To evaluate this approach, we propose a { new perplexity-based metric} that quantifies the harshness of the cut at the seam of the loop. Intuitively, if the loop boundary is truly coherent, then it should not be perceived as irregular or dissonant, neither for a human listener, nor for an audio model, as a well-trained network should roughly match human perception.

Our contributions are:
\begin{itemize}
    \item \textbf{Loop-Aware Generation via Audio Tiling}: We propose a new inference procedure that can be applied to a NAR music transformer, such as \magnet{}, to create seamlessly loopable audio samples. We call this method, and the resulting model, \texttt{LoopGen}.
    \item \textbf{Perplexity-Based Seamlessness Metric}: We introduce a metric to quantify the quality of loop boundaries, retrieving the entropy in the ``seam'' region of a track.
    \item \textbf{Empirical Validation and Code Release}: We show that our system yields superior results according to both quantitative metrics and human listening tests, and we release our code to foster future research on the generation of musical loops.
\end{itemize}

\section{Related work} 
\label{sec:rel-work}
\label{sec:related_work}

Recent advances in \textbf{music generation} leverage large-scale transformer-based architectures, which have displaced traditional recurrent neural networks for long-range sequence modeling. Pioneering systems like \emph{MuseNet}~\cite{musenet} and \emph{Music Transformer}~\cite{musictransformer} showed that attention-based models \cite{attention} could capture rich compositional structure in symbolic formats. More recently, state-of-the-art audio transformer models such as \emph{MusicGen}~\cite{musicgen} and \cite{jukebox, musiclm, melnet, llark}, have demonstrated high-quality generation of waveforms, capable of handling minutes-long clips conditioned on text or user-provided melodies.

Another wave of expressive and accurate models has come with the advent of diffusion models \cite{diffusion}  such as \emph{AudioLDM}~\cite{audioldm} and \cite{mousai, riffusion, diffariff, stableaudio, mariani2024multisource}. Their application has also reached audio and music and, in this, they are giving high quality results on-par with the transformer models.

Parallel decoding has emerged as a promising alternative to speed up generation. \emph{MAGNeT}~\cite{magnet} employs a single non-autoregressive transformer, such as those used in NLP tasks \cite{bert, elmer}, to predict masked audio tokens iteratively, showing that a well-designed masking and rescoring strategy can close the quality gap with autoregressive baselines at a fraction of the inference cost. \emph{VampNet}~\cite{vampnet}, another non-autoregressive approach, introduces inpainting capabilities and partial rewriting to refine music segments, including short repeated “vamps,” demonstrating promise for loop-centric workflows. Likewise, \emph{SoundStorm}~\cite{soundstorm} applies a bidirectional transformer on semantic tokens for efficient speech and music synthesis, further illustrating the viability of non-autoregressive methods for audio.

\textbf{Loopable music} remains comparatively underexplored. \emph{LoopNet}~\cite{loopnet} specifically targets the generation of seamless music loops, but it is tied to a limited dataset of loops which falls short of the general-purpose of free-form approaches. Other work has focused on \emph{symbolic loops} in MIDI \cite{controllable-music-loops,symbolic-music-loop}, proposing architectures that ensures segments are musically consistent when repeated. However, these methods are intrinsically different from raw audio tokens; MIDI loops require explicit pitch and instrument representations, which do not transfer to audio-generation tasks. Recently, \emph{DITTO}~\cite{dittoLoops} introduced an inference-time optimization that allows fine-grained control, including looping, over text-to-music diffusion models. While \emph{DITTO} is notable for its high output quality, it requires memory comparable to a full fine-tuning, and it slows down inference by a factor of $\sim100\times$.
% its sampling procedure incurs substantial computational cost, reportedly over 80 seconds per sample.

Finally, loopable \emph{media} generation is being tackled in computer vision with tiling techniques. Models like \emph{TileGAN}~\cite{tilegan} and \cite{seamlessgan, tileddiffusion} synthesize textures or images that repeat edge-to-edge without seams. While these visual approaches share the overarching idea of boundary alignment, they do not directly address audio continuity or musical structure. 

In this paper, we build on \emph{MAGNeT}'s non-autoregressive design to propose an inference-time approach for loopable music generation, avoiding additional training or data requirements. By treating time in a “circular” manner, our method enforces continuity at the loop boundary, substantially improving perceptual seamlessness in raw audio.

\section{Background} 
\label{sec:modeling}

\subsection{MAGNeT's inference} Unlike typical NAR models, \magnet{}'s inference does not emit all output tokens in a single inference pass. Instead, it develops the audio clip iteratively. In particular, at each iteration, \magnet{}:
\begin{enumerate}
    \item Generates logits for each empty token in the sequence.
    \item Samples a value for each token.
    \item Selects the highest scoring tokens, and marks them as fixed.
    \item Re-empties the remaining non-fixed tokens and, if no empty tokens are left, terminates; otherwise, it starts the next iteration.
\end{enumerate}
Following \cite{magnet}, we use \magnet{}'s own logits to select the tokens for the first iteration.
\magnet{}'s inference can be viewed as a generalization of autoregressive inference: rather than receiving a continuous sequence of tokens and outputting the next token, \magnet{} operates on a set of empty and non-empty tokens, filling multiple empty positions in each iteration. Thanks to its non-causal self-attention, \magnet{} can condition its outputs on both past \emph{and} future tokens, ensuring coherent generation across boundaries. This property makes \magnet{} (and similar non-autoregressive models) well-suited for loop creation, because the model can naturally attend to the loop’s start while generating its end, thereby facilitating a smoother, more seamless transition.

\subsection{Rescoring}\label{sec:rescoring}
\magnet{} \cite{magnet} also proposes a variant that linearly interpolates the probabilities given by its own logits, with those of another audio model, such as \musicgen{} \cite{musicgen}, to calculate the scores for the selection procedure. This results in a trade-off between higher quality and increased computational cost, as calculating another model's probabilities requires running it alongside \magnet{}.

\subsection{Hybrid MAGNeT}
As noted in \cite{magnet}, when presented with short audio tracks, \magnet{} produces continuations which, on average, sound better than samples created from scratch. Knowing this, we test both generating samples from scratch, and continuations of clips produced with \musicgen{}.

\section{Method}
\label{sec:method}
While our framework is, in principle, applicable to \emph{any} NAR model that generates audio iteratively, we choose \magnet{}~\cite{magnet} as our base system because it is currently the state-of-the-art in NAR music generation.

We adapt \magnet{}'s iterative inference to create a ``circular'' context around the central segment of tokens that will form our final loop. By replicating partial portions of this loop segment at the beginning and end of the generation window, \magnet{} can attend to the loop’s start when predicting its end, and vice versa. We refer to the central segment as the \emph{main loop tile}.

\subsection{Iterative overview}
\magnet{} generates audio tokens in several iterations. Each iteration partially fills an overall generation window of length \(L\). We isolate a specific subrange of length \(c\) near the center of this window to become our main loop tile. The remaining space on the left and right is filled with copies of the tile's end or beginning, respectively, thus forming a circular context.

\subsection{Inference algorithm}\label{sec:tiled_generation}
\begin{figure}[t]
    \centering
    \includegraphics[alt={Schematic showing multiple steps of the inference algorithm. At each iteration, the central tile generates tokens, which are copied into the left or right padding in the next step. Over time, the full sequence is filled, and both padding areas end up mirroring the edges of the central tile. This iterative process reinforces loop consistency.},width=\linewidth]{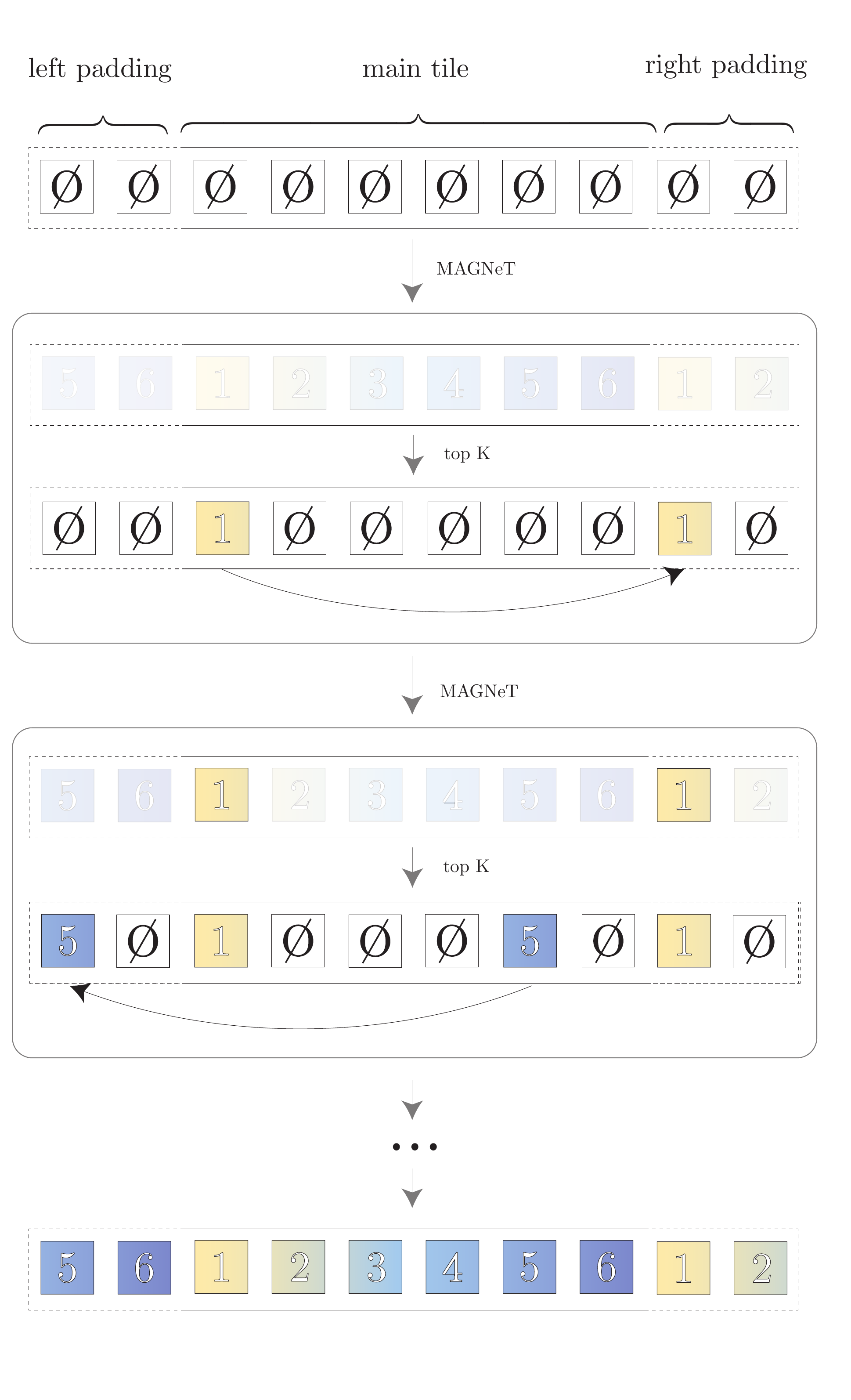}
    \vspace{-1cm}
    \caption{Diagram of our approach. The central \emph{main tile} represents the final audio segment to be looped. At each inference step, only the top-$k$ samples are maintained and reflected in the tiles. This circular padding lets \magnet{} attend to both the start and end of the tile simultaneously, ensuring a smooth transition at the loop boundary.}
    \label{fig:inference}
\end{figure}

    % \textbf{1a. Initialization.}
    At \textbf{initialization}, we start with an empty (or partially filled) window of length \(L\). In the middle of this window, we mark out \(c\) consecutive positions as the main loop tile.

    \textbf{1. Filling the Context.}
    Before calling \magnet{}, and before each inference step, we copy:
    \begin{itemize}
        \item The \emph{ending} of the main tile into the \emph{left} side of the window, so that the first tokens of the tile can ``see'' what happens at the end of it.
        \item The \emph{beginning} of the main tile into the \emph{right} side of the window, so that the last tokens of the tile can ``see'' its start.
    \end{itemize}
    This ensures a fully circular arrangement: the model effectively observes how the loop's end meets its beginning.

    \textbf{2. MAGNeT Inference.}
    We run \magnet{} on the entire window of length \(L\). Because \magnet{} uses non-causal (bidirectional) attention, tokens in the main tile can be conditioned on both the left-side copy (its own end) and the right-side copy (its own start).

    \textbf{3. Token Selection.}
    At the end of each iteration, only tokens \emph{within the main tile} are considered for finalizing. We keep those that \magnet{} assigns the highest probability (e.g., top-$k$ or threshold-based), marking them as fixed (i.e., no longer empty in subsequent iterations). The rest are reset to empty.

    \textbf{4. Repeat until completion.}
    We move on to the next inference iteration, going back to step \textbf{2}, until the entirety of the \emph{main tile} is filled.

    \textbf{5. Extract the final loop.}
    Once the iteration limit is reached or all main-tile tokens are fixed, the algorithm stops. The central \(c\) tokens (our main loop tile) are extracted as the final result. Repeating this tile end-to-start yields a seamless loop.

\subsection{Hybrid variant}
\label{hybrid-variant}
\magnet{} often produces higher-quality audio when continuing from a given prompt rather than generating entirely from scratch \cite{magnet}. To take advantage of this, we first generate an audio segment ${C}$ with \musicgen{}, empirically set to half the desired final clip length. For instance, if the final clip is intended to last $10$s, we let \musicgen{} produce the first $5$s and then provide these tokens as a partially filled {main tile}. This approach forces the model to generate a coherent continuation of the high-quality prompt, ensuring the ending transitions seamlessly to the beginning. Empirically, we observe that this {hybrid} version surpasses samples generated without an audio prompt in terms of semantic variety, objective audio quality, and musical coherence.

\subsection{Signature-aware length control}
\label{sec:bpm-signature}

A well-formed loop often sounds most musical when it aligns with full bars (e.g., 2 or 4 bars of consistent tempo). Generating loops of arbitrary length may create awkward breaks if, for instance, the tempo does not fit integer bar divisions.

To mitigate this, we use the current state-of-the-art beat-extraction system, \texttt{beat\_this} \cite{beatthis} on the initial audio prompt \({C}\), to identify:
\begin{itemize}
    \item The average duration between beats, \(\delta \approx 60/\mathrm{BPM}\).
    \item The median number of beats per bar, \(\mu\).
\end{itemize}
We use these to compute the duration of a bar that the prompt $C$ implies. We want the duration of the entire loop $l$ to be \emph{a)} an exact multiple (or submultiple) of the duration of a bar; \emph{b)} constrained in a time interval \([\alpha, \beta]\). To achieve this, we repeatedly double or halve the initial candidate length $l$ until it fits these constraints.
% Then we compute a candidate bar length $l$, which we fit within user-specified bounds \([\alpha, \beta]\), while ensuring it remains a musically coherent unit, by repeatedly doubling or halving it, and finally, we run the tiled-generation procedure with this chosen \(C\).

\begin{algorithm}
    \begin{algorithmic}
        \Require Audio clip $C$, min/max duration $\alpha$ and $\beta$, preferred number of bars $n$
        \State $B, D\gets$ detected beats/downbeats in \texttt{beat\_this}$(C)$ 
        \State $\delta\gets$ median time elapsed between $B$s \Comment{akin to $\frac{60}{\textrm{BPM}}$}
        \State $\mu\gets$ median \#beats between $D$s \Comment{bar of the clip}
        \State $l\gets n\mu\delta$ \Comment{duration of $n$ bars}
        \While{$l<\alpha\vee l>\beta$}
        \If{$l < \alpha$}
        \State $l \gets 2l$
        \Else
        \State $l \gets \frac{l}{2}$
        \EndIf
        \EndWhile
        \If{$l\in [\frac{n\mu\delta}{4}, 4n\mu\delta]$} \Comment{Too far from $n$ bars?}
        \State \Return $l$ \Comment{Return if sufficiently close}
        \Else
        \State \Return $\varnothing$ \Comment{Otherwise abort (try another $C$)}
        \EndIf
    \end{algorithmic}
    \caption{Beat Alignment algorithm}\label{alg:beat-alignment}
\end{algorithm}

When used in combination, tiled generation and the beat alignment \Cref{alg:beat-alignment} produce loops that not only have smooth seam transitions but also respect musical structure. This results in clips that are more naturally \emph{loopable} for applications like music production, live performance, or any setting in which tightly aligned repeating segments are required.

\section{Evaluation Metrics}
\label{sec:good-music-loop}
When assessing looped music, a clip that sounds fine in a single pass may still have an abrupt transition when it repeats. Standard metrics such as FAD \cite{kilgour2019frechetaudiodistancemetric}, which measure the overall distributional similarity between generated and real audio using neural embeddings, may miss such artifacts. Because FAD operates over full-length clips, it can overlook localized issues such as discontinuities at the loop boundary. To address this, we propose a metric that focuses directly on the transition seam.

\subsection{Seam perplexity}
To automatically assess the continuity of the loop around the seam, we adapt the idea of \emph{perplexity} from language modeling. 
Let us assume that we have a well-trained music generation model (such as \musicgen{}) that can estimate probabilities for each token (or audio frame) in a music clip. While traditional perplexity sums over all tokens in a clip, we focus exclusively on the seam, that is, the transition point where the end meets the beginning, where loop artifacts are most likely to occur.

\begin{figure}
     \centering
     \includegraphics[alt={Line plot comparing cross-entropy of audio tokens around a loop point for several generation methods. Most methods show a sharp spike in cross-entropy at the loop seam, indicating poor loop quality, before quickly decreasing. LoopGen, the proposed method, avoids this spike entirely.},width=\linewidth]{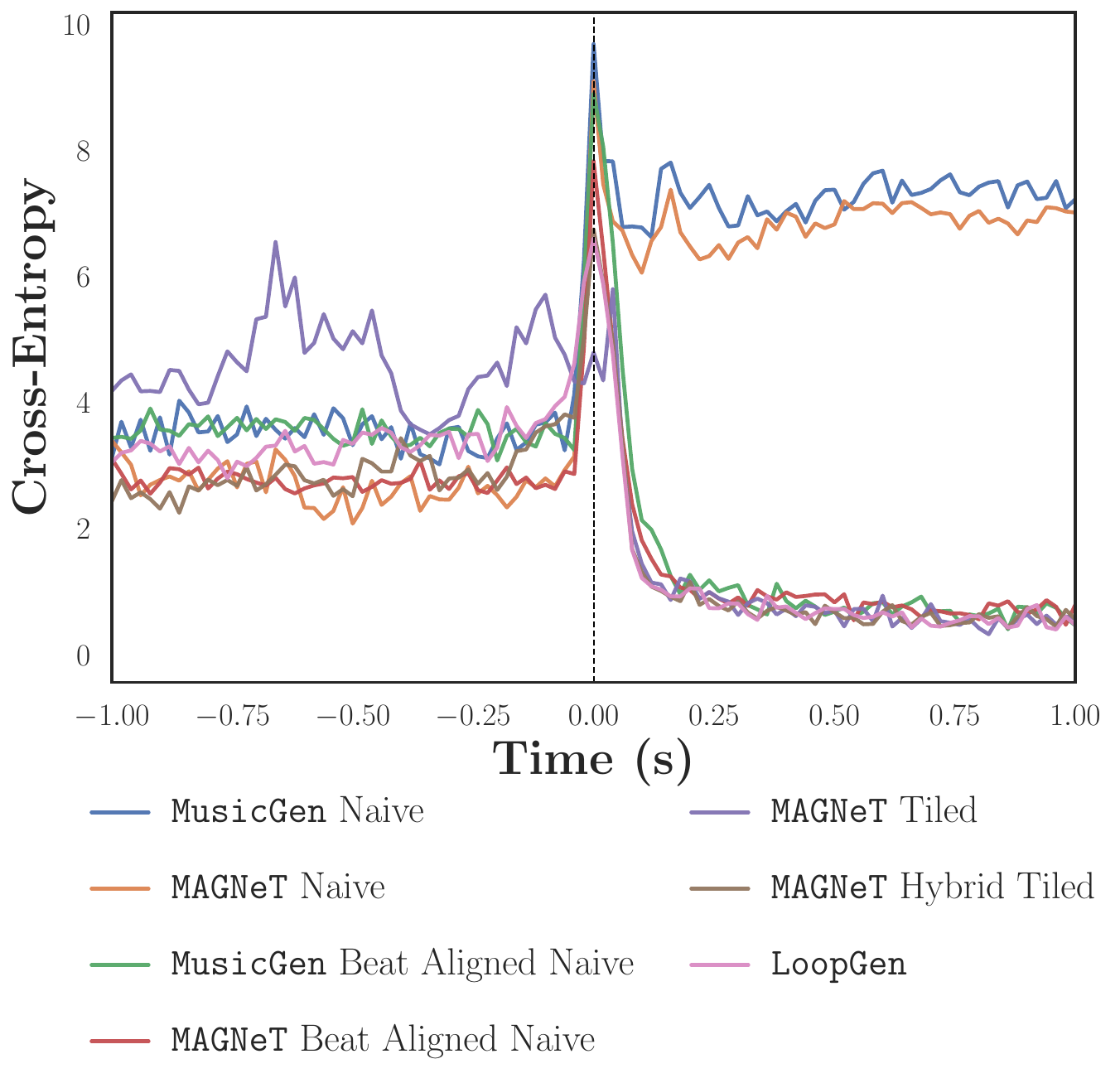}
     \caption{Average cross entropy of \musicgen\ around the seam (highlighted with a dashed line) for different models/variants.}
     \label{fig:avg-cross-entropy}
\end{figure}

\subsubsection{Cross-entropy and perplexity}
First recall that, for a sequence \(X = (x_1, x_2, \ldots, x_T)\), the average cross-entropy \(H(X)\) is:
\begin{equation}
H(X) = -\frac{1}{T} \sum_{i=1}^T \ln \mathcal{M}(x_i).
\end{equation}
Intuitively, if \(\mathcal{M}\) assigns higher probability to each token, the cross-entropy will be smaller, indicating better alignment between model and data. From cross-entropy, we derive the perplexity \(\mathcal{P}(X)\), a standard measure of how well a model predicts a sequence:
\begin{equation}
\mathcal{P}(X) = \exp\bigl(H(X)\bigr) 
               = \exp\!\Bigl(-\frac{1}{T}\sum_{i=1}^T\ln\mathcal{M}(x_i)\Bigr).
\end{equation}
A lower perplexity value indicates that \(\mathcal{M}\) finds the sequence more predictable (or more likely).

\subsubsection{Seam perplexity}
While global perplexity focuses on the entire clip, loop artifacts, as in \Cref{fig:avg-cross-entropy}, occur specifically at the boundary where the clip wraps around. To isolate how the model perceives that transition, we compute \emph{seam perplexity} on a short window around the boundary.

Let us be given \(N\) generated clips \(\{X^{(k)}\}\). Each \(X^{(k)}\) has length \(T\), and we identify a seam boundary at index \(b^{(k)}\). We then define a window of size \(W\) immediately following \(b^{(k)}\), i.e., the tokens
\begin{equation}
X^{(k)}_{\text{seam}} = \bigl(x_i^{(k)} : i \in [b^{(k)},\,b^{(k)} + W - 1]\bigr).
\end{equation}
The average cross-entropy of the seam tokens in \(X^{(k)}\) is:
\begin{equation}\label{eq:seam_centropy}
H_{\text{seam}}(X^{(k)}) 
= -\frac{1}{W} \sum_{i = b^{(k)}}^{b^{(k)} + W - 1} 
     \ln\mathcal{M}\bigl(x_i^{(k)}\bigr).
\end{equation}
Finally, the {seam perplexity} is the exponential of the mean seam cross-entropy across all \(N\) clips:
\begin{equation}
\text{Seam Perplexity} 
= \exp\!\Bigl(\tfrac{1}{N}\sum_{k=1}^N H_{\text{seam}}\bigl(X^{(k)}\bigr)\Bigr) \,.
\end{equation}

A low seam perplexity indicates that the seam is ``easy'' for a strong reference model to predict, suggesting a smooth transition. Conversely, a high value suggests abrupt discontinuities or other artifacts at the loop boundary.

\section{Experiments}
\label{sec:experiments}
In the following, all generated tracks are conditioned with the same set of $100$ textual prompts, and \magnet{}'s iterations are set to $\langle100, 50, 10, 10\rangle$ for each of the 4 codebooks from \encodec{} \cite{encodec} respectively. Textual prompts are generated automatically via a LLM, some of them are (e.g.):

\ex. ``\emph{A high-energy EDM track with a powerful drop and sidechain compression}''

\ex. ``\emph{An Irish folk dance tune with energetic fiddle and bodhrán drum}''

\subsection{Evaluating models}
\subsubsection{Baselines} For both \magnet{} and \musicgen{}, two baseline solutions are formulated: (i) \textbf{Naive}, a sample is generated and repeated, without further processing, and (ii) \textbf{Beat-Aligned (BA) Naive}, a sample is generated, ran through \Cref{alg:beat-alignment} to cut them at a musically-valid length, and repeated.
\subsubsection{Our techniques} From the contributions of this paper, three models are evaluated: (i) \textbf{Tiled}, samples are generated via the tiled generation technique described in \Cref{sec:tiled_generation}, (ii) \textbf{Hybrid Tiled}, samples are generated with the same tiled technique, but starting from an audio prompt generated by \musicgen{} (ref. \Cref{hybrid-variant}), and (iii) \textbf{Beat Aligned Tiled}, which uses the same technique as \textbf{Hybrid Tiled}, but with the additional application of the beat-alignment algorithm described in \Cref{sec:bpm-signature}. This latter variant is our best performing model, and what, going forward, we call \texttt{LoopGen}.
For each variant, both Seam Perplexity and \texttt{fadtk}'s \cite{fadtk} FAD (with embeddings from both VGG-ish\cite{vggish} and CLAP \cite{laionclap} over the FMA-Pop \cite{fma_dataset} dataset) are computed.

\subsection{Hyperparameter search}\label{sec:hyperparameter_search}
The most important hyperparameters for the samples' quality we identify are {classifier-free guidance} ($\lambda$) and the {rescoring coefficient} ($\omega$). The former controls how much a model should adhere to the conditioning information given (in our case, the textual prompt), instead of following the emerging sample. In \magnet{}'s original paper, the authors find that the best FAD is reached with $\lambda=10.0$ (linearly decreasing to $\lambda=1.0$ as the iterations pass) but, as the tiling constraint might increase the contextual information that the model can gain from the input, we verified that a lower coefficient translates into more organic generations.

The rescoring coefficient $\omega$, instead, controls the interpolation coefficient introduced in \Cref{sec:rescoring}. When $\omega=0$, rescoring is not applied, when $1$, only \musicgen{}'s probabilities are used. We test therefore our algorithm with multiple coefficients ranging from 0 to 1.
A reasonable value for the cfg was chosen to be $\lambda=5.0$; on the other hand, the rescoring was chosen through a thorough search conducted on both \magnet{} \textbf{Tiled} and \magnet{} \textbf{Hybrid Tiled}, generating 100 10-seconds samples for each model. Our results, presented in \Cref{tab:hyperparam-experiments}, empirically show the best rescoring to be $\omega=0.5$.

It is worth noting that the \textbf{Hybrid} version of the model consistently achieves better FAD scores, but worse perplexity. The better FAD score can be clearly attributed to the initial prompt generated by \musicgen{}, which consistently surpasses \magnet{}'s audio quality. This hybrid combination of different models is also the reason for the increase in perplexity, since the final generation consists of a concatenation of tokens sampled from different distributions.

\begin{table}[h]
    \centering
    \resizebox{1\linewidth}{!}{%
        \begin{tabular}{c l c c c}
            \toprule
            Model     & Variant     & $\omega$ & $\textrm{FAD}_\textrm{vggish}\ (\downarrow)$ & $\textrm{Seam Perplexity}$ ($\downarrow$) \\
            % \cmidrule{1-3} \cmidrule(l){4-4}
            \midrule
            %\magnet{} & Tiled      & $10.0$ & $0.0$    & $2.92$             & $28.09\pm7.36$                         \\
            \magnet{} & Tiled       & $0.0$    & $3.05$             & $23.88\pm5.40$                         \\
            \magnet{} & Tiled       & $0.25$   & $3.22$             & $25.17\pm5.35$                         \\
            \magnet{} & Tiled        & $0.50$   & $3.51$             & $\mathbf{18.15\pm3.53}$                         \\
            \magnet{} & Tiled       & $0.75$   & $3.97$             & $24.55\pm5.43$                         \\
            \magnet{} & Tiled       & $1.0$    & $4.35$             & $25.42\pm4.86$                         \\
            \magnet{} & Hybrid Tiled  & $0.0$    & $2.97$             & $39.30\pm7.21$                         \\
            \magnet{} & Hybrid Tiled & $0.25$   & $2.99$             & $47.72\pm10.11$                        \\
            \magnet{} & Hybrid Tiled & $0.50$   & $2.98$             & $44.42\pm9.33$                         \\
            \magnet{} & Hybrid Tiled & $0.75$   & $3.00$             & $43.93\pm8.39$                         \\
            \magnet{} & Hybrid Tiled & $1.0$    & $\mathbf{2.93}$             & $41.74\pm9.05$                         \\
            \bottomrule
        \end{tabular}
    }
    \caption{Rescoring experiments ($\lambda=5.0$)}
    \label{tab:hyperparam-experiments}
    
\end{table}

\subsection{Final results}
Below, we present our final results across six baselines and our three novel models, generated with the same previous 100 textual prompts, but 30 seconds long. \textbf{Tiling} models, using the technique described in \Cref{sec:tiled_generation}, exhibit significantly lower Seam Perplexity compared to their non-tiled counterparts, though at the cost of a weaker FAD score. However, \texttt{LoopGen}, leveraging both the \textbf{Hybrid} approach (\Cref{hybrid-variant}) and \Cref{alg:beat-alignment}, achieves the best FAD score among all models. This improvement comes with a slight increase in Seam Perplexity, as previously discussed.

Despite this minor trade-off in perplexity, \texttt{LoopGen} substantially outperforms baseline solutions, offering a more musically pleasing output due to its alignment with rhythmically meaningful cut points (\Cref{alg:beat-alignment}). This results in tracks that maintain better musical coherence compared to the standard \textbf{Tiled} model.

\begin{figure}[h]
    \centering
    \includegraphics[alt={Ridgeplot showing the distribution of seam perplexity scores for several variants. The plot visualizes the data from Table 2. MAGNeT Tiled shows slightly lower perplexity than LoopGen, though their distributions are similar. In contrast, baseline methods show much higher perplexity, indicating worse performance at loop transitions.},width=\linewidth]{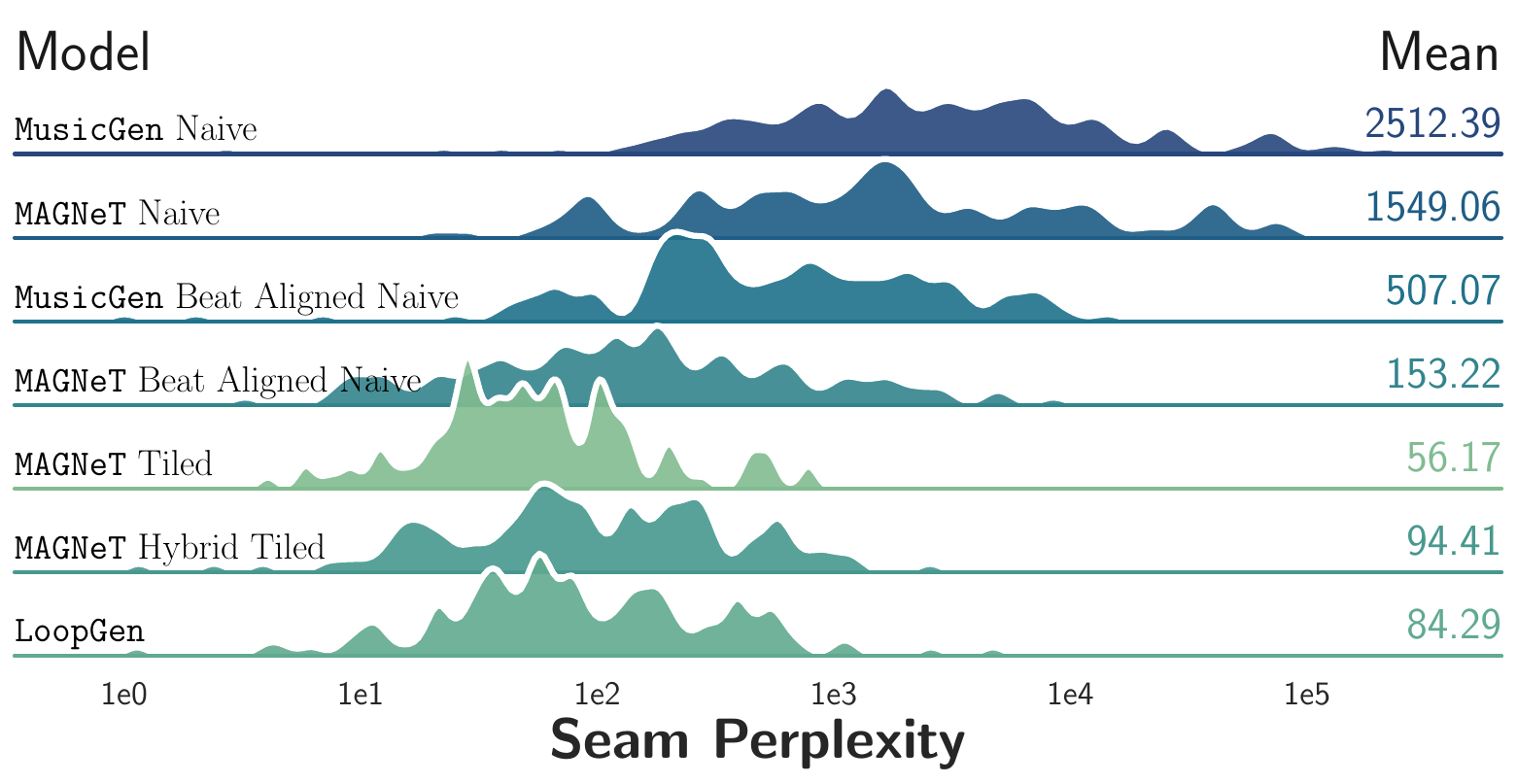}
    \caption{\textbf{Seam Perplexity} distribution of the considered models (lower is better). }
    \label{fig:seam_perplexity}
\end{figure}

\Cref{tab:main_experiments} presents the evaluation metrics, and the distribution of Seam Perplexity values is visualized in \Cref{fig:seam_perplexity}.
\begin{table}[h]
    \centering
    \resizebox{1\linewidth}{!}{%
        \begin{tabular}{c l c c c}
            \toprule
            Model       & Variant       & $\textrm{FAD}_\textrm{vggish}\ (\downarrow)$ & $\textrm{FAD}_\textrm{CLAP}\ (\downarrow)$ & $\textrm{Seam Perplexity}$ ($\downarrow$) \\
            % \cmidrule{1-3} \cmidrule(l){4-4}
            \midrule
            \magnet{}   & Vanilla       & $3.36$   & $0.33$          & ---                                    \\
            \magnet{}   & Naive    & $3.36$      & $0.35$       & $1549.06\pm556.03$                       \\
            \magnet{}   & Beat Aligned Naive & $3.34$   & $0.34$          & $153.22\pm47.69$                       \\
            \cdashlinelr{1-5}
            \musicgen{} & Vanilla       & $2.81$  & $0.32$           & ---                                    \\
            \musicgen{} & Naive    & $2.81$   & $0.33$          & $2512.39\pm903.16$                      \\
            \musicgen{} & Beat Aligned Naive & $2.86$     & $0.33$        & $507.07\pm163.67$                       \\
            \cdashlinelr{1-5}
            \magnet{} & Tiled       & $4.30$     & $0.51$        & $\mathbf{56.17\pm11.78}$                         \\
            \magnet{} & Hybrid Tiled & $2.98$     & $0.33$        & $94.41\pm25.77$                         \\
            \magnet{} & \texttt{LoopGen} & $\mathbf{2.80}$      & $\mathbf{0.31}$       & $84.29\pm22.66$                        \\
            \bottomrule
        \end{tabular}
    }
    \caption{Main experiments' evaluation metrics. For each model, we compute the FAD with VGG-ish and CLAP embeddings using FMA-pop as a reference dataset. For reference, we also compute FAD scores for both \magnet{} and \musicgen{}'s standard, non-looping, generations).}
    \label{tab:main_experiments}
\end{table}

\subsection{Human evaluation}
Using the previous setups, we prepare a set of: 100 10-seconds clips from \texttt{LoopGen} (ours), and  100 10-seconds clips from \magnet{} Hybrid Naive (without Tiling-generation, baseline).
We select the latter model because it is the most similar to ours, without any of the modifications introduced in this paper. This ensures a fair comparison, with the primary expected difference being seamlessness. The clips are chosen to be 10 seconds long for ease of listening.

With this set of samples, we conduct a blind listening experiment with a group of users. Each volunteer listens to up to 30 randomly selected clips (15 from our model, 15 from the baseline) and rates the perceptibility of the seam on a Likert scale (1 = Evident cut, 5 = Imperceptible cut).
\begin{figure}[h]
     \centering
     \includegraphics[alt={Histogram comparing the distribution of listening test rankings for our technique versus the baseline. Our method shows a strong skew toward higher rankings, with its mean near rank 4. The baseline is skewed toward lower rankings, averaging around rank 2.},width=\linewidth]{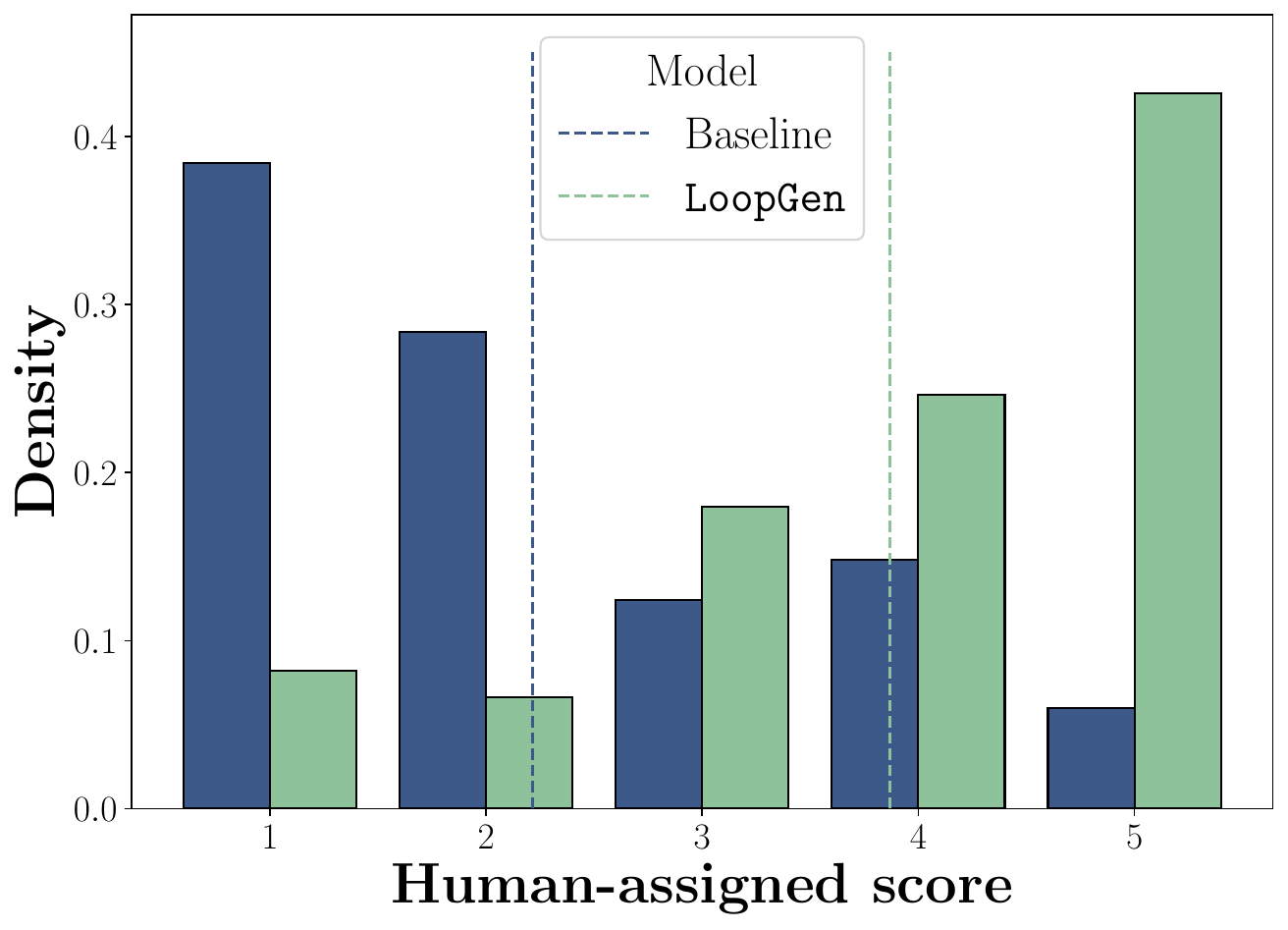}
     \caption{Distribution of perceptibility ratings, comparing \texttt{LoopGen} with the baseline. Lines are model's mean.}
     \label{fig:human-eval}
\end{figure}
In total, we collect $506$ data points $s$ from $18$ listening sessions.
Computing each user's average rating of each model; we run a paired t-test such that 
\begin{equation}
H_0\equiv \mu_\textrm{\texttt{LoopGen}}=\mu_\textrm{baseline}
\end{equation}
This yields $t(17)=12.21,\, p<10^{-9}$, providing overwhelming evidence against $H_0$. Furthermore, the effect size is large ($d=2.88$), confirming a very strong evidence that our technique substantially reduces the perceptibility of the seam, as can also be seen in \Cref{fig:human-eval}.

\section{Conclusions} 
\label{sec:conclusions}
With this research, we have introduced a novel inference-only approach for generating loopable music, leveraging a simple “circular” padding scheme within MAGNeT’s non-autoregressive framework to ensure seamless boundaries. Our experiments demonstrated clear gains in loop continuity, validated both by a new perplexity-based seam metric and by human listening tests. The whole procedure does not require additional training or specialized loop datasets. By aligning loop length to musical beats, the generated audio segments more naturally fit common compositional structures, further improving their usability in practice. Overall, this work underscores the potential of lightweight, inference-time solutions for enhancing generative music models.

\section{Acknowledgments}
This work is supported by Sapienza University of Rome via the Seed of ERC grant ``MINT.AI'', cup B83C25001040001. Furthermore, we thank all of the participants in the human evaluation test.

\bibliography{main}

\newpage

\onecolumn
\appendix
\section{Seam Perplexity's Error Margins}
In various tables, we present the values of our Seam Perplexity as $\textrm{center}\pm\textrm{standard error}$. Because perplexity is the exponentiation of average cross-entropy, it is impossible to actually compute error margins directly. To obtain these values, we start from \Cref{eq:seam_centropy} and compute for each dataset of samples $\mathbf{X}=\{X^1,\dots,X^N\}$ the mean cross-entropy:
\begin{equation}
    \mu_\mathbf{X} = \tfrac{1}{N}\sum_{k=1}^N H_{\text{seam}}\bigl(X^{(k)}\bigr)
\end{equation}
and standard deviation
\begin{equation}
    \sigma_\mathbf{X} = \sqrt{\frac{1}{N-1}\sum_{k=1}^N\left(\mu_\mathbf{X} -  H_{\text{seam}}\bigl(X^{(k)}\bigr)\right)^2}.
\end{equation}
We then compute the 95\% confidence intervals for the cross-entropy
\begin{equation}
    \left[\mu_\mathbf{X}-1.96\tfrac{\sigma_\mathbf{X}}{\sqrt{N}},\mu_\mathbf{X}+1.96\tfrac{\sigma_\mathbf{X}}{\sqrt{N}}\right],
\end{equation}
and transform them into exponential space
\begin{equation}
        \left[l=\exp\left(\mu_\mathbf{X}-1.96\tfrac{\sigma_\mathbf{X}}{\sqrt{N}}\right),r=\exp\left(\mu_\mathbf{X}+1.96\tfrac{\sigma_\mathbf{X}}{\sqrt{N}}\right)\right].
\end{equation}
Finally, we calculate the provided values as
\begin{equation}
    \textrm{center}=\tfrac{1}{2}(l+r),\qquad \textrm{standard error} = \tfrac{1}{2}(r-l).
\end{equation}
This approach differs from the common method of showing a value with error margins, where the error is modeled as Gaussian, and the center value is assumed to be the empirical mean of the measured quantity. In this case, however, since the perplexity operation itself is computed as the exponentiation of its mean, it would be impossible to calculate a symmetric Gaussian error margin directly (not without running calculations on multiple folds of the data).

\section{10 seconds experiments}
\begin{figure}[h]
    \centering
    \includegraphics[alt={Ridgeplot showing the distribution of seam perplexity scores for various audio generation methods on 10-second samples (compared to 30-second samples in the main paper). The data corresponds to Table 3. MAGNeT Tiled again shows slightly better perplexity than LoopGen, while all baseline methods have much higher perplexity, indicating poorer loop quality.},width=.7\linewidth]{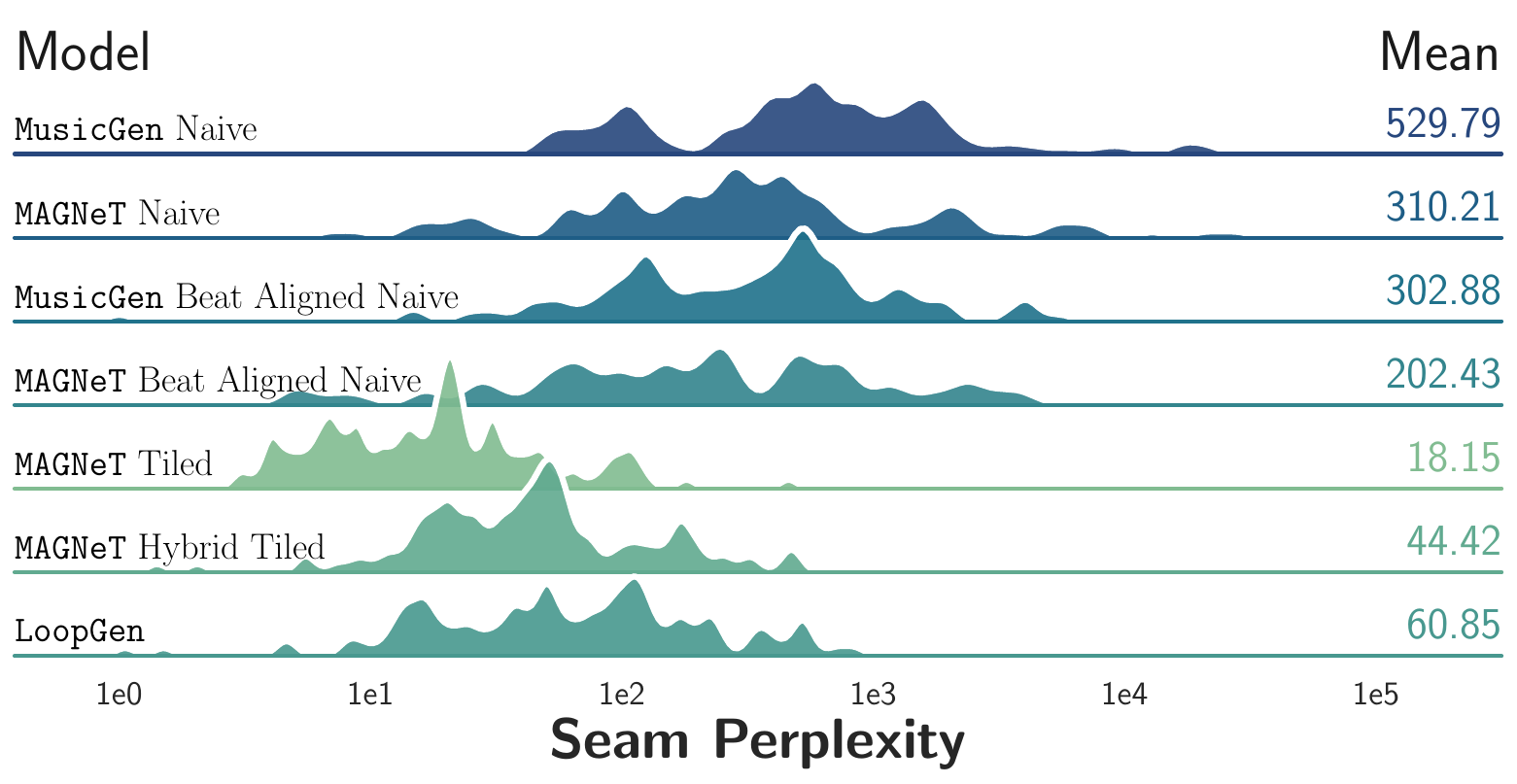}
    \caption{\textbf{Seam Perplexity}'s distributions for 10 seconds samples (lower is better).}
    \label{fig:ten_seconds_seam_perplexity}
\end{figure}

During development, we also explored the same final experiments seen in the main article (\Cref{tab:main_experiments}) with the 10 seconds variant of \magnet{}. The results of these experiments are detailed in \Cref{tab:ten_seconds_experiments} and visualized in \Cref{fig:ten_seconds_seam_perplexity}. Notably, the \textbf{Seam Perplexity} exhibits a significant change with this modification.  While it is unclear whether this change is solely attributable to the different models, the shorter track length, or a combination thereof, we empirically observed no discernible perceptual difference in the seamlessness of the 10-second and 30-second samples.
\begin{table}[h]
    \centering
    \resizebox{.7\linewidth}{!}{%
        \begin{tabular}{c l c c c}
            \toprule
            Model       & Variant  & $\textrm{FAD}_\textrm{vggish} (\downarrow)$ & $\textrm{FAD}_\textrm{CLAP} (\downarrow)$ & $\textrm{Seam Perplexity}$ ($\downarrow$) \\
            % \cmidrule{1-3} \cmidrule(l){4-4}
            \midrule
            \magnet{}   & Vanilla       & $3.05$    & $0.39$          & ---                                    \\
            \magnet{}   & Naive    & $3.02$       & $0.31$      & $310.21\pm98.33$                       \\
            \magnet{}   & Beat Aligned Naive & $3.03$      & $0.35$      & $202.43\pm67.23$                       \\
            \cdashlinelr{1-4}
            \musicgen{} & Vanilla       & $3.28$    & $0.41$      & ---                                    \\
            \musicgen{} & Naive    & $3.21$        & $0.34$     & $529.79\pm167.87$                      \\
            \musicgen{} & Beat Aligned Naive & $3.24$         & $0.31$     & $302.88\pm79.91$                       \\
            \cdashlinelr{1-4}
            \magnet{} & Tiled       & $3.51$       & $0.40$      & $18.15\pm3.53$                         \\
            \magnet{} & Hybrid Tiled & $2.98$      & $0.33$       & $44.42\pm9.33$                         \\
            \magnet{} & \texttt{LoopGen} & $2.95$          & $0.33$     & $60.85\pm15.24$    \\
            \bottomrule
        \end{tabular}
    }
    \caption{10 seconds versions of main experiments' evaluation}
    \label{tab:ten_seconds_experiments}
\end{table}
\newpage

\section{Beat alignment algorithm visualization}
\begin{figure}[h]
    \centering
    \includegraphics[alt={Diagram showing a waveform with clearly visible peaks representing beats. The first few beats are labeled as if detected by a beat detection algorithm. Based on this beat timing, the beat alignment algorithm determines the optimal point to cut the audio sample, ensuring musical continuity.},width=.5\linewidth]{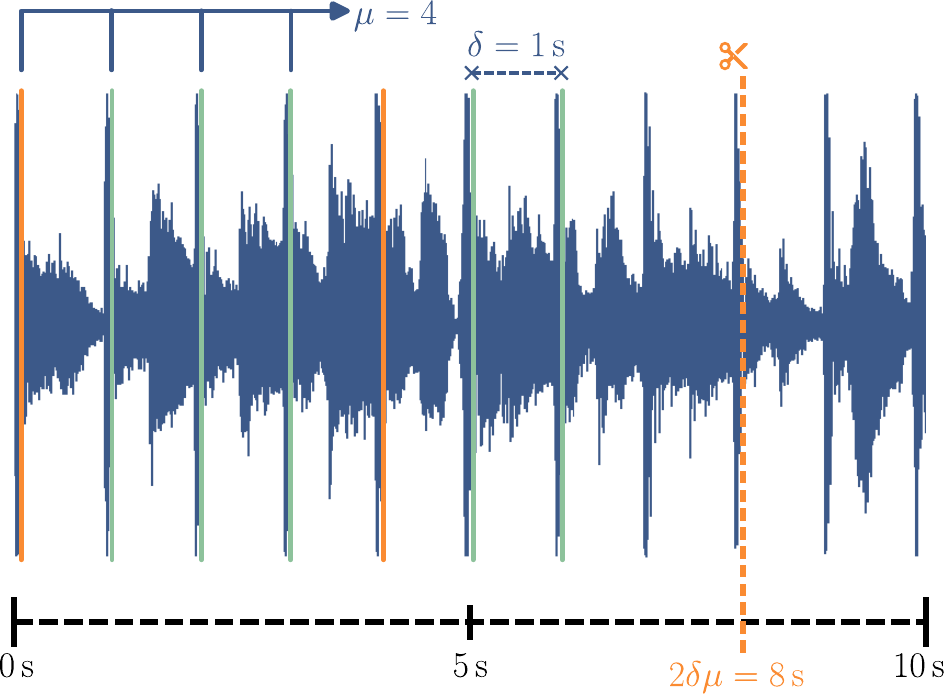}
    \caption{Visualization of our beat alignment algorithm (\Cref{alg:beat-alignment}). Downbeats are shown in orange, and beats in light green. If the waveform were cut at the 10-second mark, the resulting unit would be 2.5 bars long—musically unpleasing when repeated. Instead, our algorithm ensures cuts occur at a whole-bar length or, if not possible, at a multiple.}
    \label{fig:beat_alignment_visualization}
\end{figure}

To further illustrate the inner workings of our beat alignment algorithm, we provide a visual representation in Figure \ref{fig:beat_alignment_visualization}. This figure highlights the key aspects of the algorithm as applied to a typical waveform. Specifically, it showcases how the algorithm ensures that audio segments are cut at musically meaningful locations, preventing awkward segment lengths that could disrupt the perception of rhythmic continuity.

\section{MAGNeT's Inference}
In the original paper of \magnet{} \cite{magnet}, the authors provide a detailed pseudocode of the inference procedure, we show the same in \Cref{fig:modified-procedure}, with our added lines highlighted. It can be seen how our modification is actually pretty agnostic to the model behaviour and, it can be argued, that it may indeed be applied to other models which act in a similar manner.

\begin{figure}[h]
    \centering
    \begin{lstlisting}[language=Python, style=mystyle, caption=]
def loopgen_MAGNeT_generate(B: int, T: int, c: int, text: List, s: int, model: nn.Module,
                    rescorer: nn.Module, mask_id: int, tempr: float, w: float):
    
    # Start from a fully masked sequence
    gen_seq = torch.full((B, T), mask_id, dtype=torch.long)
    
    n_spans = T #// span_len    To fine-control which samples are picked
    spans_shape = (B, n_spans)
    span_scores = torch.zeros(spans_shape, dtype=torch.float32)

    <@\textcolor{additioncolor}{\# Addition 1. Calculate limits of main tile}@>
    l = min((T - c) // 2, c)
    r = T - c - l
    
    # Run MAGNeT iterative decoding for 's' iterations
    for i in range(s):
        mask_p = torch.cos((math.pi * i) / (2 * s))
        n_masked_spans = max(int(mask_p * n_spans), 1)
        
        # Masking
        masked_spans = span_scores.topk(n_masked_spans, dim=-1).indices
        mask = get_mask(spans_shape, masked_spans)
        gen_seq[mask] = mask_id

        <@\textcolor{additioncolor}{\# Addition 2. Copy main tile to padding areas}@>
        if r > 0:
            gen_seq[..., :l] = gen_seq[..., -l-r:-r]
            gen_seq[..., -r:] = torch.cat([gen_seq[..., l:-r]] * (r // c + 1), -1)[..., :r]
        
        # Forward pass
        logits, probs = model.compute_predictions(gen_sequence, text, cfg=True, temperature=tempr)
        
        # Classifier free guidance with annealing
        cfg_logits = cfg(mask_p, logits, annealing=True)
        
        # Sampling
        sampled_tokens = sample_top_p(probs, p=top_p)
        
        # Place the sampled tokens in the masked positions
        mask = gen_seq == mask_id
        gen_seq = place_sampled_tokens(mask, sampled_tokens[..., 0], gen_seq)
        
        # Probs of sampled tokens
        sampled_probs = get_sampled_probs(probs, sampled_tokens)
        if rescorer:
            # Rescoring
            rescorer_logits, rescorer_probs = rescorer.compute_predictions(gen_seq, text)
            rescorer_sampled_probs = get_sampled_probs(rescorer_probs, sampled_tokens)
           
            # Final probs are the convex combination of probs and rescorer_probs
            sampled_probs = w * rescorer_sampled_probs + (1 - w) * sampled_probs

        <@\textcolor{additioncolor}{\# Addition 3. Give no probability to padding areas}@>
        sampled_probs[..., :l] = 0
        sampled_probs[..., -r:] = 0
        
        # Span scoring - max
        span_scores = get_spans_scores(sampled_probs)
        
        # Prevent remasking by placing -inf scores for unmasked
        span_scores = span_scores.masked_fill(~spans_mask, -1e5)
    
    return gen_seq
\end{lstlisting}
    \caption{Modified \magnet{} inference procedure, stylized for ease of reading (wouldn't actually run correctly in its current form), our added lines are highlighted in light-brown.}
    \label{fig:modified-procedure}
\end{figure}

\end{document}